\documentclass[prl,preprint,superscriptaddress, letterpaper, fleqn, floatfix, showpacs]{revtex4}
\usepackage{euscript, units, amsfonts,amsmath, graphics, graphicx, dcolumn, fancyhdr}
\usepackage{times}
\usepackage{sidecap}

\usepackage[usenames]{color}

\begin{document}

\title{Self-Diffusion in 2D Dusty Plasma Liquids: Numerical Simulation Results}

\author{Lu-Jing Hou}
\affiliation{IEAP, Christian-Albrechts Universit\"{a}t zu Kiel,
D-24098 Kiel, Germany}
\author{Alexander Piel}
\affiliation{IEAP, Christian-Albrechts Universit\"{a}t zu Kiel,
D-24098 Kiel, Germany}
\author{P. K. Shukla}
\affiliation{Institut f\"{u}r Theoretische Physik IV, Ruhr-Universit\"{a}t Bochum, D-44780, Germany}
\begin{abstract}
We perform Brownian dynamics simulations for studying the self-diffusion in two-dimensional (2D) dusty
plasma liquids, in terms of both mean-square displacement and velocity autocorrelation function (VAF).
Super-diffusion of charged dust particles has been observed to be most significant at infinitely small
damping rate $\gamma$ for intermediate coupling strength, where the long-time asymptotic behavior
of VAF is found to be the product of $t^{-1}$ and $\exp{(-\gamma t)}$. The former represents the
prediction of early theories in 2D simple liquids and the latter the VAF of a free Brownian particle.
This leads to a smooth transition from super-diffusion to normal diffusion, and then to sub-diffusion
with an increase of the damping rate. These results well explain the seemingly contradictory scattered
in recent classical molecular dynamics simulations and experiments of dusty plasmas.
\end{abstract}

\received{3 December 2008}

\maketitle

Diffusion processes play an important role in determining dynamical properties of many physical,
chemical and biological systems \cite{Hansen}. Diffusion in two-dimensional (2D) physical systems is
of particular interest because of its significance in fundamental physics \cite{Alder1970,Ernst1970,Dorfman1970},
as well as its applications in many 2D systems, such as strongly-correlated electrons on the liquid helium
surface \cite{Grimes1979}, colloidal suspensions \cite{Murry1990}, core-softened fluid \cite {CSF},
strongly coupled non-neutral plasmas \cite{Mitchell1999} and, in particular, strongly coupled dusty
plasmas (SCDPs) \cite{Shukla2002}, which have been generally recognized as a promising model system
to study many phenomena in solids, liquids and other strongly coupled Coulomb systems at the kinetic
level and have been used frequently to study many physical processes in 2D systems,
such as heat conduction \cite{Heat}, shear flow \cite{Shear} and certainly diffusion of
charged dust particles \cite{Juan1998,Quinn2002,Ratynskaia2005,Ratynskaia2006,Bin2006,Bin2007,Bin2008,
Ott2008,Nunomura2006,Vaulina2006,Donko2009}.

Diffusion is usually described by using particles' mean-square displacement (MSD) and velocity autocorrelation
function (VAF), defined by $\text{MSD}(t)=\langle \vert \mathbf{r}_{i}(t)-\mathbf{r}_{i}(0) \vert^2 \rangle$
and $Z(t)=\langle \mathbf{v}_{i}(t)\cdot\mathbf{v}_{i}(0) \rangle/\langle v^2_{i}(0)\rangle$, respectively.
Here $\mathbf{r}_{i}(t)$ and $\mathbf{v}_{i}(t)$ are the position and velocity of $i$th particle at time $t$,
respectively, and the angular bracket denotes an ensemble average. Note that the (normal) diffusion coefficient
$D$ is determined by the MSD through the Einstein relation $D=\lim_{t\rightarrow\infty}\text{MSD}(t)/(4t)$
and by VAF through Green-Kubo formula $D=\int_{0}^{\infty}{Z(t)dt}/(2\Gamma)$ \cite{Hansen}. What makes
things interesting in 2D systems probably lies in the fact that the diffusion processes could exhibit some
anomalous behaviors. For the normal diffusion, one would expect $\text{MSD}(t)\propto t$, and $\vert Z(t)\vert$
decays faster than $t^{-1}$, when $t\rightarrow\infty$, in order to obtain a meaningful diffusion coefficient.
Diffusion that do not comply with these characteristics are usually regarded as anomalous one, for example
the so-called super-diffusion and sub-diffusion. In the former case (super-diffusion), one
expects \cite{Alder1970,Ernst1970,Dorfman1970} $\text{MSD}(t)\propto t^{1+\alpha}$, where $\alpha$ is
positive, and $\vert Z(t)\vert\propto t^{-1}$, when $t\rightarrow\infty$ and in the latter (sub-diffusion),
$\alpha$ becomes negative, when $t\rightarrow\infty$.

Early molecular dynamics (MD) simulations with 2D hard-disk liquids \cite{Alder1970} observed an asymptotic
decay of the VAF in the form that is $\propto t^{-1}$, which led to the conclusion that particle motions
in 2D hard-disk systems are non-diffusive (or super-diffusive). Later theories of 2D simple
liquids \cite{Ernst1970,Dorfman1970} came to the same conclusion, but with a wider argument saying that
this asymptotic behavior should also exist in kinetic parts of correlation functions for the shear
viscosity and heat conductivity, regardless of the liquid density and interaction force, and finally
lead to the breakdown of the hydrodynamics in 2D systems.

The anomalous diffusion of dust particles in quasi-2D and 2D dusty plasma liquids (DPLs) has been frequently
reported in many laboratory experiments \cite{Juan1998,Quinn2002,Ratynskaia2005,Ratynskaia2006,Bin2008}.
In particular, the super-diffusion of charged dust grains in 2D (monolayer) experiments has been recently observed
by different groups \cite{Bin2008,Quinn2002,Ratynskaia2005,Ratynskaia2006}. Quinn and Goree \cite{Quinn2002}
were the first to  report the observation of super-diffusion at long timescales in a 2D DPL experiment,
followed by Ratynskaia \emph{et al.} \cite{Ratynskaia2005,Ratynskaia2006}, and more recently by Liu and
Goree \cite{Bin2006,Bin2007,Bin2008}. These observations were supported by the early theoretical
prediction \cite{Alder1970,Ernst1970,Dorfman1970} and recent MD simulations with 2D and quasi-2D
Yukawa liquids \cite{Bin2006,Bin2008,Ott2008,Donko2009}.

However, seemingly contradictory observations exist in both experiments and simulations.
Nunomura \emph{et al.} \cite{Nunomura2006} investigated the self-diffusion in 2D DPLs for a wide
range of the Coulomb coupling strength, and observed a normal diffusion process till the melting point,
at which charged dust particles are frozen and therefore a sub-diffusion was observed. Their observation
was supported by their own MD simulation results, and recent Brownian dynamics simulation\cite{Vaulina2006},
in which the neutral damping effect was considered and normal diffusion was observed in the whole liquid region.
Yet, 2D simulations with softer interactions \cite{CSF} observed normal diffusive motion.

The present work is particularly motivated by these contrasting observations in recent 2D dusty plasma experiments
and simulations, and is intended to clarify some of the above incompatibility, for example, the effects of the neutral
gas damping and stiffness of the dust particle interaction on super-diffusion, by using extensive numerical
simulations.

Our simulation is based on the Brownian dynamics (BD) method \cite{Allen1989}, which brings the simulation closer
to real dusty plasma experiments, as the effects of the neutral gas damping and dust particle Brownian motions
are included in a self-consistent manner. Most relevant experiments are those conducted in equilibrium
conditions \cite{Quinn2002,Nunomura2006,Ratynskaia2005}.

In our simulation, $N=20000$ particles are placed in a square with periodical boundary conditions (PBC).
A 5$th$-order Gear-like Predictor-Corrector algorithm \cite{Hou2008pop} was used to integrate the Langevin
equations of every interacting Brownian dust particles. As usual, we assume here pairwise Yukawa interaction
between charged dust particles: $\Phi(r)=(Q^2/r)\exp{(-r/\lambda_{D})}$, where $Q$ is the charge on a
dust particle, $r$ is the interparticle distance, and $\lambda_{D}$ is the Debye screening length.
Such a system can be fully characterized by three parameters \cite{Fortov2003}: the Coulomb coupling
parameter $\Gamma=Q^2/(aT)$, the screening parameter $\kappa=a/\lambda$, and the damping rate
$\gamma/\omega_\mathrm{pd}$ due to the neutral gas, where $T$ is the system temperature (in energy units),
$a=(\pi n)^{-1/2}$ the Wigner-Seitz radius, and $n$ the equilibrium dust number density.
The dusty plasma frequency is $\omega_\mathrm{pd}=\left[2Q^{2}/(m a^{3})\right]^{1/2}$, where $m$ is the dust
particle mass. The system size is approximately $250a \times 250a$, which together with PBC and the
dust-acoustic speed \cite{Kalman2004} decides our maximum independent observation time to be
$320 \omega^{-1}_\mathrm{pd}$.

\begin{figure}[htp]
\centering
\includegraphics[trim=12mm 5mm 20mm 10mm,clip, width=0.6\textwidth]{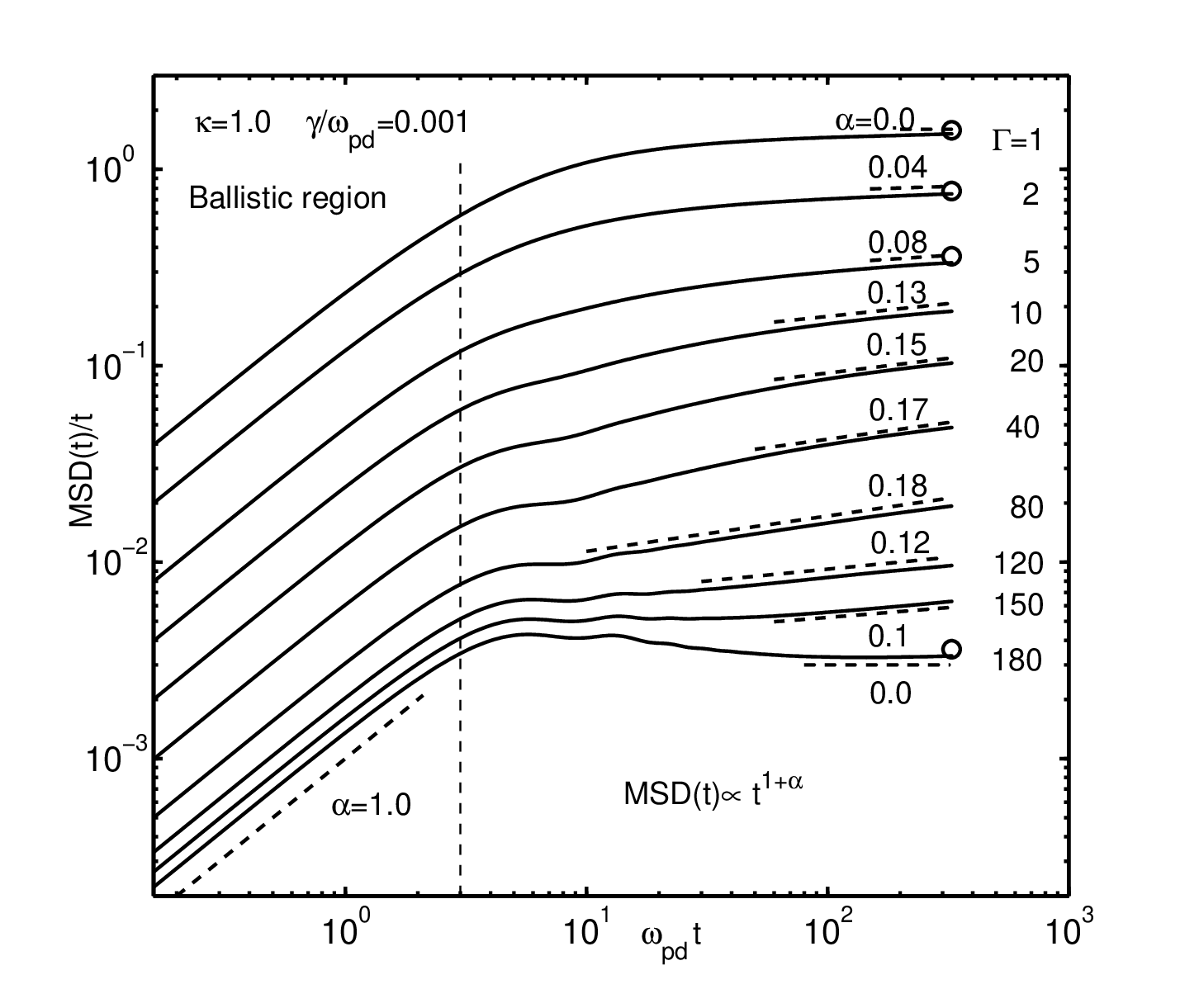}
\caption{$\text{MSD}(t)/t$ (normalized by $\omega_\mathrm{pd}a^2$) for $\kappa=1.0$,
$\gamma/\omega_\mathrm{pd}=0.001$ and different $\Gamma$. Dash-lines are linear fits (in log-log scale)
of asymptotic behaviors at long time limit, $\alpha$ is the slope of the fits. Lengths of the dash-lines
show roughly the fitting range.  Typical uncertainty is $\pm 8\%$ of the shown values. Circles are
corresponding diffusion coefficients obtained through Green-Kubo formula. }
\label{Fig_MSD_k1_g0001}
\end{figure}

\begin{figure}[htp]
\centering
\includegraphics[trim=8mm 5mm 22mm 11mm,clip, width=0.6\textwidth]{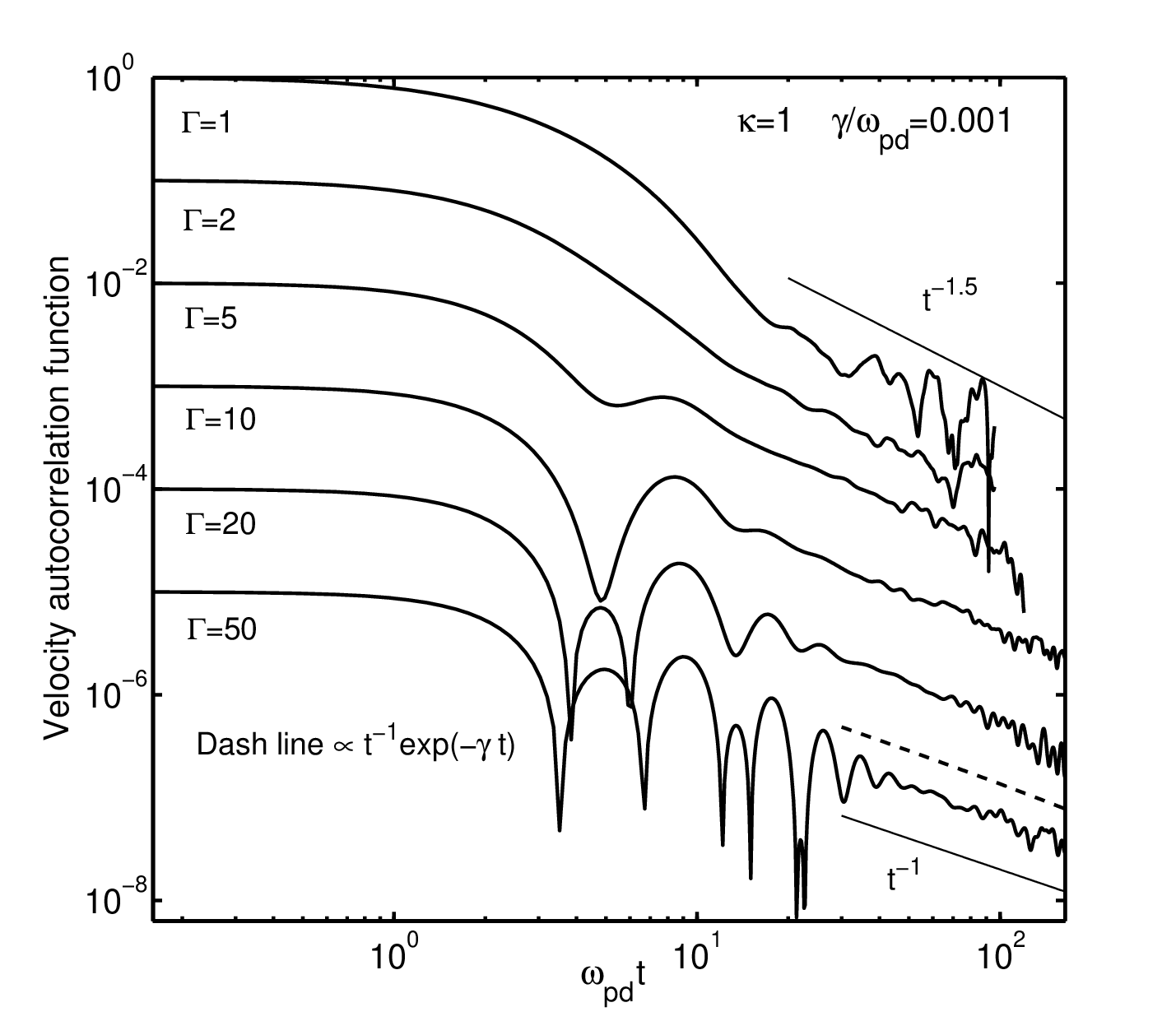}
\caption{The velocity autocorrelation function for $\kappa=1.0$, $\gamma/\omega_\mathrm{pd}=0.001$
and different $\Gamma$.  The decay of $t^{-1}$ and $t^{-1.5}$ are also shown for comparison.
Dash-line follows $t^{-1}\exp{(-\gamma t)}$.} \label{Fig_VAF_k1_g0001}
\end{figure}

Our analysis will be based on MSD and VAF. For a better visualization of different types of diffusion processes, we have used throughout this paper the
quantity $\text{MSD}(t)/t$, rather than $\text{MSD}(t)$ itself. Let us firstly see the evolution of MSD in
the zero damping limit. Figure \ref{Fig_MSD_k1_g0001} displays $\text{MSD}(t)/t$ for $\kappa=1.0$ and
$\gamma/\omega_\mathrm{pd}=0.001$, but for different $\Gamma$, which covers from nonideal
gaseous state ($\Gamma=1$) to the melting point ($\Gamma=180$) \cite{Kalman2004}. Shown together in this figure
are linear fits (in log-log scale) of their asymptotic behaviors at short and long time limits, respectively.
The former gives a constant value of $\alpha=1$, indicating the ballistic motion, whereas the latter varies
for different $\Gamma$, as depicted by the values above or below the corresponding dash lines. The variation
of $\alpha$ with $\Gamma$ at long time limit shows clearly a hill shape, with a flat peak of $\alpha$ in the range
of $\Gamma\in [50, 80]$ and a peak value of $\alpha\approx 0.18$. This behavior indicates a clear transition
from normal diffusion $\rightarrow$ super-diffusion $\rightarrow$ normal diffusion, when $\Gamma$ goes from $1$
to $180$. Super-diffusion is most significant at the intermediate Coulomb coupling strength. The general tendency
agrees qualitatively with previous MD simulations of Ott \emph{et al.} (in the 2D limit) \cite{Ott2008} and
Liu and Goree \cite{Bin2007}, except that in the latter the peak appears around $\Gamma=20$ with a value of
$\alpha \approx 0.3$. This discrepancy could have been caused by the limited observation timescale
(which then was determined by the finite size of the system.) in Ref. \cite{Bin2007}. Their latest
simulations \cite{Bin2008} with substantially more particles gave a peak value very close to ours. In addition,
the diffusion coefficients calculated by using the Green-Kubo formula were also shown as circles for $\Gamma$
with a normal diffusive behavior. Their horizontal location indicates the integration limits. We found a
very close agreement on the diffusion coefficient by using these two methods, which proves the consistence
of our simulation.
\begin{figure}[htp]
\centering
\includegraphics[trim=11mm 8mm 15mm 10mm,clip, width=0.6\textwidth]{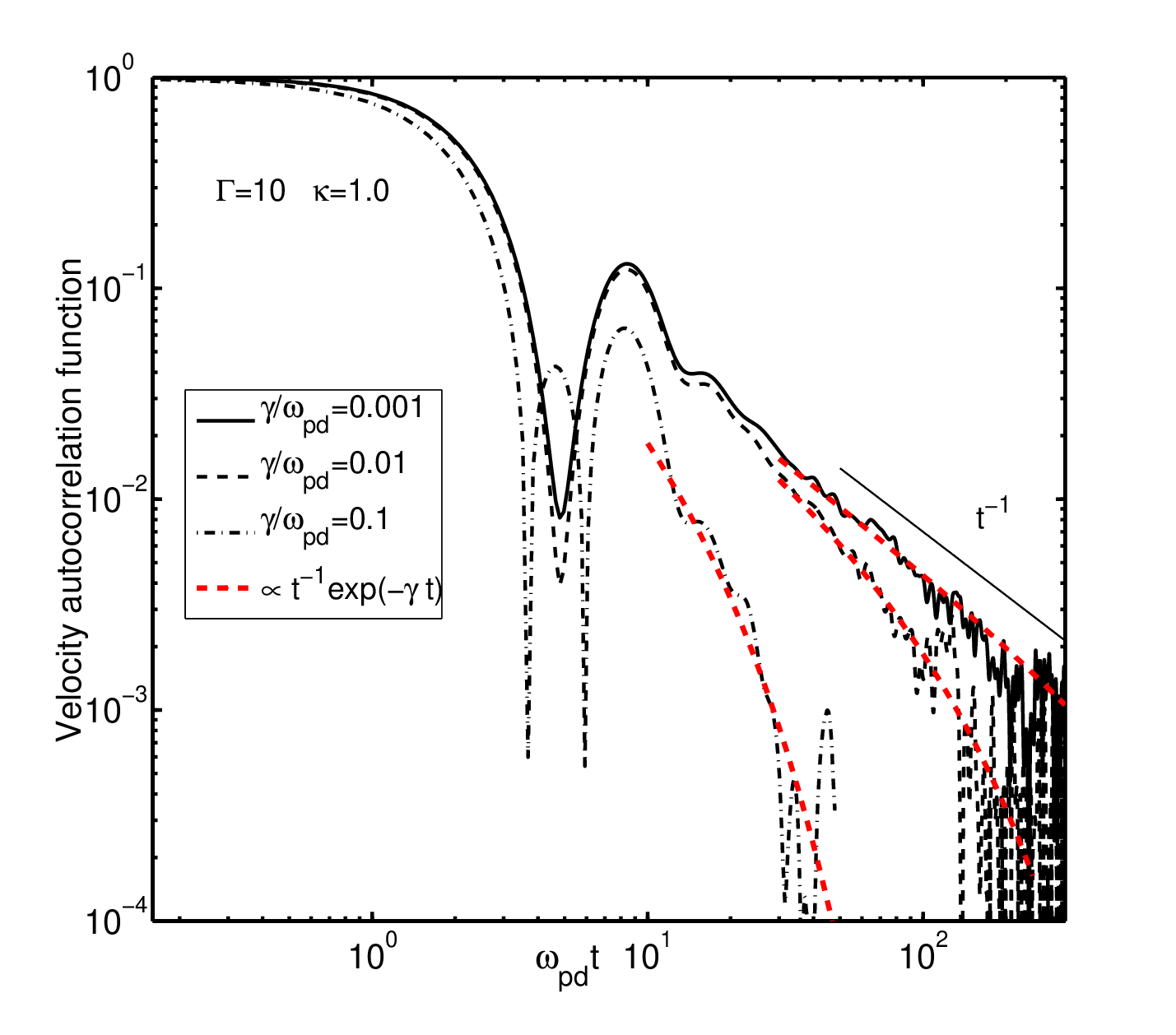}\caption{The velocity autocorrelation
function for $\kappa=1.0$, $\Gamma=10$, and different damping rate $\gamma$ (normalized by $\omega_\mathrm{pd}$).
Heavy dash-lines follow $t^{-1}\exp{(-\gamma t)}$. The decay of $t^{-1}$ is also shown for comparison. }
\label{Fig_VAF_k1_G_10}
\end{figure}

\begin{figure}[htp]
\centering
\includegraphics[trim=15mm 5mm 10mm 0mm,clip, width=0.7\textwidth]{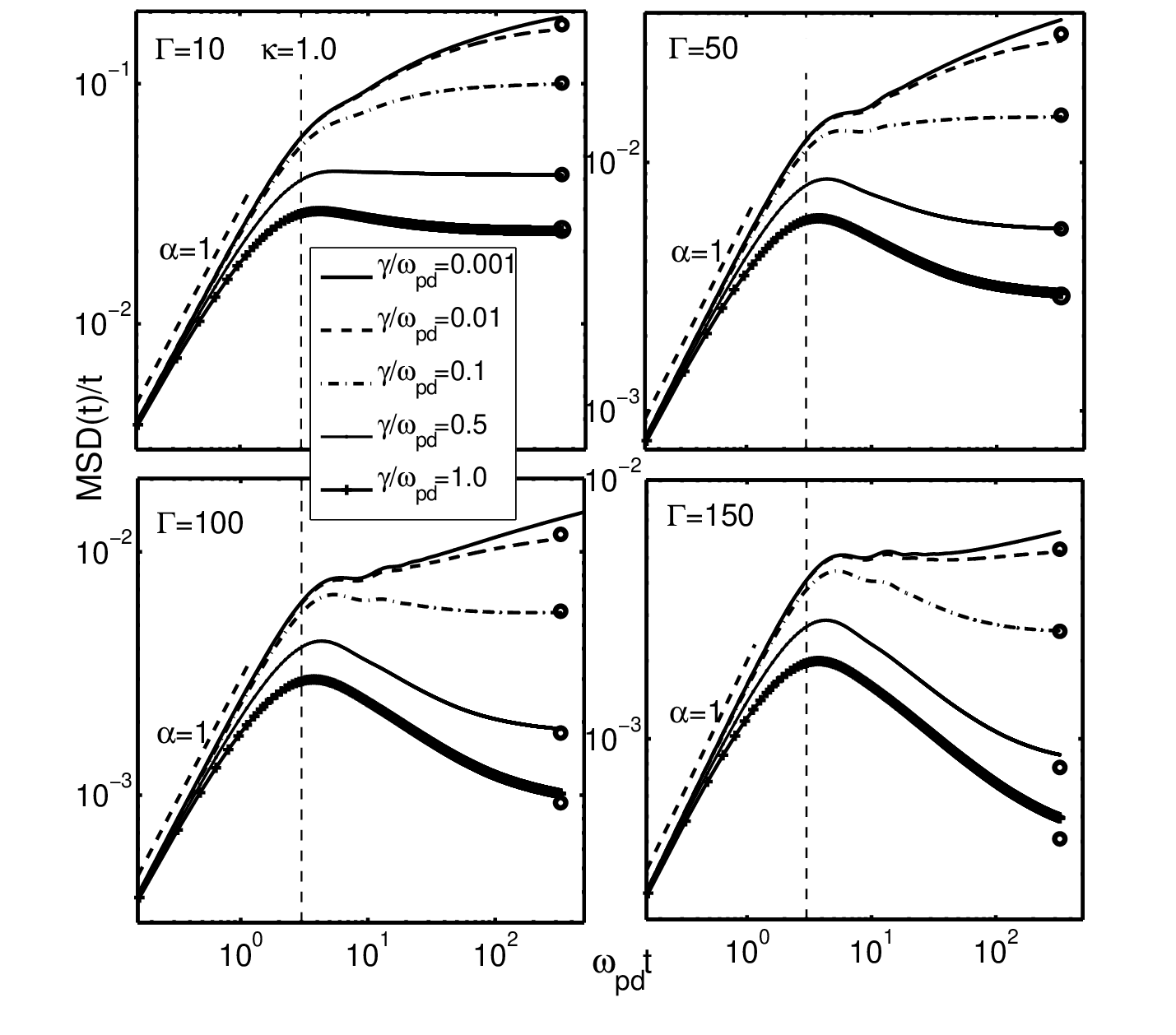}
\caption{$\text{MSD}(t)/t$ (normalized by $\omega_\mathrm{pd}a^2$) for $\kappa=1.0$, but different $\Gamma$
and damping rates (normalized by $\omega_\mathrm{pd}$). Circles are corresponding to diffusion coefficients
obtained through the Green-Kubo formula. } \label{Fig_MSD_k1}
\end{figure}

Figure \ref{Fig_VAF_k1_g0001} depicts the VAFs (their absolutes) for different $\Gamma$ in the same conditions
as Fig. \ref{Fig_MSD_k1_g0001}. Shown together in this figure are different decay rates following $t^{-1.5}$, $t^{-1}$
and $t^{-1}\exp{(-\gamma t)}$, respectively. One observes a continuous transition from a fast decay at $\Gamma=1$
(faster than $t^{-1.5}$) to a slow decay at $\Gamma=50$ approximately following $t^{-1}$, indicating the occurrence
of the super-diffusion. The last three tails all follow roughly $t^{-1}$ decay, however, they seem to be better
fitted by $t^{-1}\exp{(-\gamma t)}$, as shown by dash lines in the figure, with the exponential part coming from
the VAF of a free Brownian dust particle. Since the exponential factor obviously suppresses the super-diffusion,
what we observe here is a super-diffusion in a transient timescale. In other words, there should be no absolute
super-diffusion (in a sense of infinitely long time limit) in a damped system in the equilibrium state.
However, for this specific damping rate $\gamma/\omega_\mathrm{pd}=0.001$, its effect is negligibly small
at our maximum timescale. One would need a timescale of order $10^{3}\omega^{-1}_\mathrm{pd}$, which then
require about one million dust particles, to observe a complete suppression of the super-diffusion in this case.
Nevertheless, this tendency will be clearer next when we vary the damping rate.

Figure \ref{Fig_VAF_k1_G_10} displays the VAF for $\kappa=1.0$ and $\Gamma=10$ but different $\gamma$, as a
typical example exhibiting the damping effect on the long time tails of the VAF. The long time decay of all VAFs
can be well fitted in the form of $t^{-1}\exp{(-\gamma t)}$ up to about $\gamma/\omega_\mathrm{pd}=0.1$. For higher
damping rates, the VAF reaches the noise level (about $10^{-3}$) too fast, so that no reliable fits can be applied.
It should also be noted that a damping as small as $\gamma/\omega_\mathrm{pd}=0.01$ is sufficient to bring the
decay close to $t^{-1.5}$ in our timescale and to suppress the super-diffusion. Since the typical value in similar
dusty plasma experiments \cite{Nunomura2006,Ratynskaia2005,Bin2008} is about a few percent of $\omega_\mathrm{pd}$,
we would expect that there should be no super-diffusion in the equilibrium state. This explains well the
absence of the super-diffusion in the experiment of Nunomura \emph{et al.} \cite{Nunomura2006}.

Figure \ref{Fig_MSD_k1} exhibits the variation of $\text{MSD}(t)/t$ for $\kappa=1.0$ but for different
$\Gamma$ and $\gamma$ values. When $\gamma$ goes from infinitely small to a finite value, another transition
from the super-diffusion $\rightarrow$ normal diffusion $\rightarrow$ sub-diffusion is clearly seen at long
timescale especially from panels with higher $\Gamma$, which implies an enhancement of the caging effect due
to the neutral damping. This explains the sub-diffusion close to the melting point observed in the experiment of
Nunomura \emph{et al.} \cite{Nunomura2006}.

Circles in Fig. \ref{Fig_MSD_k1} correspond to the diffusion coefficients $D$ obtained through the
Green-Kubo formula, which agree with those from MSD (last points of $\text{MSD}(t)/t$) very well. Relative
big discrepancies are found for the highest Coulomb coupling strength $\Gamma=150$ and highest damping
rates $\gamma/\omega_\mathrm{pd}=0.5$ and $1.0$, which implies that our timescale is not long enough to
reach the steady value.
\begin{figure}[htp]
\centering
\includegraphics[trim=18mm 8mm 22mm 12mm,clip, width=0.6\textwidth]{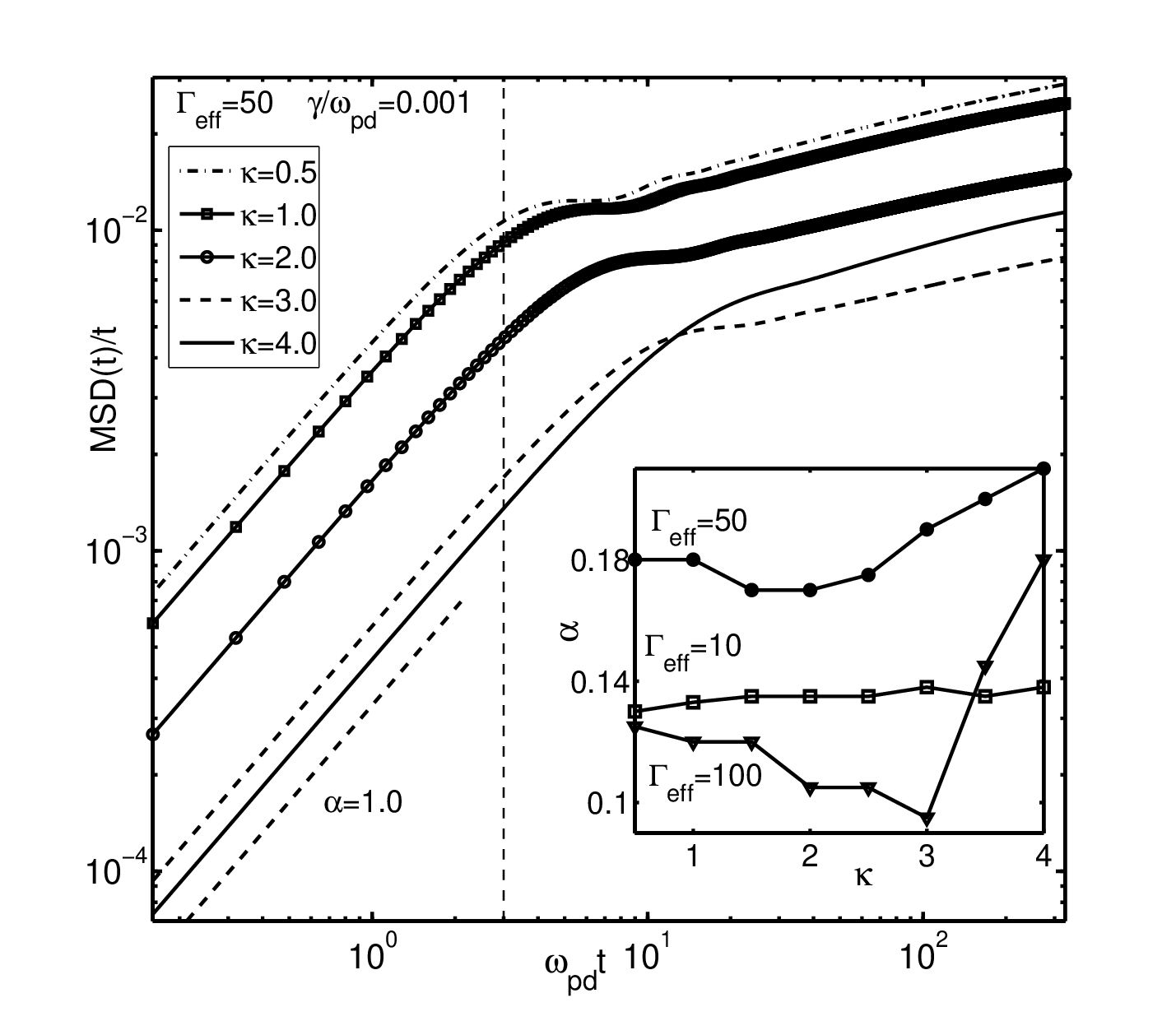}\caption{$\text{MSD}(t)/t$
(normalized by $\omega_\mathrm{pd}a^2$) for $\Gamma_\mathrm{eff}=f(\kappa)\Gamma=50$ ($f(\kappa)$ is given in
Ref. \cite{Kalman2004}), $\gamma/\omega_\mathrm{pd}=0.001$ and different $\kappa$. The inserted figure shows
exponential $\alpha$ obtained by fitting at long time limit for different $\Gamma_\mathrm{eff}$,
with other conditions the same. The uncertainties in the fits are $\pm 13\%$ of the shown values
for $\Gamma_\mathrm{eff}=10$ and $\pm 8\%$ for $\Gamma_\mathrm{eff}=50$ and $100$, respectively.}
\label{Fig_MSD_G50}
\end{figure}

One sees from both circles and long time behavior of $\text{MSD}(t)/t$ that $D$ decreases with the increase
of the damping rate. This is physically clear because the damping effect retards the dynamics of charged dust
particles. This tendency also agrees qualitatively with observations in 3D BD
simulations \cite{Vaulina2002,Vaulina2006} and in experiments \cite{Fortov2003} of dusty plasma liquids,
in which $D$ was found to decay as $1/(1+\gamma/\omega_\mathrm{pd})$ with the damping rate, in a certain range of
the Coulomb coupling parameter \cite{Vaulina2002,Fortov2003,Vaulina2006}.

Finally, we have investigated the effect of stiffness of dust particle interactions on the diffusion processes.
Here we vary the screening parameter $\kappa$, while keeping the effective Coulomb coupling
parameter $\Gamma_\mathrm{eff}=f(\kappa)\Gamma$ constant, where the universal scaling function $f(\kappa)$ is obtained
by fitting the MD simulation results \cite{Kalman2004}. Figure \ref{Fig_MSD_G50} shows the variations of MSD
for $\Gamma_\mathrm{eff}=50$, $\gamma/\omega_\mathrm{pd}=0.001$, but for different $\kappa$ values, as examples.
It is seen that the dust particle motion at this coupling strength (and timescale, to be more precisely) is
super-diffusive.  When $\kappa$ increases, the interaction becomes stiffer and stiffer and more and more
like a hard-disk system.  During this course, one observes that the migration of dust particles becomes
weaker and weaker due to the decrease of the temperature, except at the highest $\kappa$, where the diffusion
is enhanced shortly after the ballistic regime. The effect on the super-diffusion seems complicated, as is
shown (the inserted figure) by the variation of exponential $\alpha$ at long time limit for different
$\Gamma_\mathrm{eff}$ values.  Varying $\kappa$ has little effect for $\Gamma_\mathrm{eff}=10$, as the $\alpha$
rises slightly from $0.13$ at $\kappa=0.5$ to $0.138$ at $\kappa=4.0$. For $\Gamma_\mathrm{eff}=50$ and $100$,
the largest value of $\alpha$ was observed at highest $\kappa$, indicating the most significant super-diffusion.
The observation seems contradictory with the theoretical prediction of Ernst \emph{et al.} \cite{Ernst1970},
in which they predicted that  the super-diffusion motion should exist in all 2D liquids regardless of the
interaction force, but in accordance with a previous speculation based on many computer simulations that
the super-diffusion might be more significant in systems with stiffer interactions. However, we did not
observe a monotonous increase of $\alpha$ with an increase of $\kappa$ (hence stiffness of interactions),
which indicates that dependence of the super-diffusion on the interaction force might be much more complicated
than one would have thought.

In summary, we have studied the self-diffusion processes in 2D dusty plasma liquids by using the Brownian
dynamics simulation. The latter shows, for the first time, the transition from super-diffusion to normal diffusion
then to sub-diffusion with the increase of the damping rate. In particular, our observations suggest that there
should be no real super-diffusion at long time limit in a 2D dusty plasma liquids in the equilibrium state,
such as the experimental observations by Nunomura \emph{et al.} \cite{Nunomura2006}. Most of other experiments
of super-diffusion were made in systems that were not in a strict equilibrium state and therefore collective
excitations generally cannot be ruled out. This applies to the experiments of Ratynskaia \emph{et al.}
\cite{Ratynskaia2005,Ratynskaia2006}, where super-diffusion was observed in a viscoelastic vortical dust fluid,
and to the experiments of Liu and Goree \cite{Bin2006,Bin2008}, which were made in a driven dissipative system.
Our simulation results, therefore, may stimulate further experimental investigations.

\begin{acknowledgments}
L.J.H. gratefully acknowledges the financial support from the Alexander von Humboldt foundation. The work at
CAU is supported by DFG within SFB-TR24/A2 and by DLR grant 50WM 0739.
\end{acknowledgments}

\end{document}